\documentclass[10pt,superscriptaddress,twocolumn,amsmath,amssymb,aps,prl,showpacs]{revtex4-2}
\usepackage{mathrsfs}
\usepackage{graphicx}
\usepackage{dcolumn}
\usepackage{bm}

\usepackage{graphicx}
\usepackage{epstopdf}

\usepackage{amsmath,amssymb,mathrsfs}
\usepackage{braket}
\usepackage{hyperref}
\hypersetup{
    colorlinks=true,
    linkcolor=blue,
    filecolor=blue,
    urlcolor=blue,
    citecolor=magenta
    }

\begin{document}


\title{Dynamics of strongly interacting Fermi gases with time-dependent interactions: Consequence of conformal symmetry}


\author{Jeff Maki}
\affiliation{Department of Physics and HKU-UCAS Joint Institute for Theoretical and Computational Physics at Hong Kong, The University of Hong Kong, Hong Kong, China}
\author{Shizhong Zhang}
\affiliation{Department of Physics and HKU-UCAS Joint Institute for Theoretical and Computational Physics at Hong Kong, The University of Hong Kong, Hong Kong, China}
\author{Fei Zhou}
\affiliation{2) Department of Physics and Astronomy, University of British Columbia, 6224 Agricultural Road, Vancouver, BC, V6T 1Z1, Canada} 

\date{\today}

\begin{abstract} 
In this Letter, we investigate the effects of a time-dependent, short-ranged interaction on the long-time expansion dynamics of Fermi gases. We show that the effects of the interaction on the dynamics is dictated by how it changes under a conformal transformation, and derive an explicit criterion for the relevancy of time-dependent interactions in both the strongly and non-interacting nearly scale invariant quantum gases. In addition, we show that it is possible to engineer interactions that give rise to non-exponential thermalization dynamics in trapped Fermi gases. To supplement the symmetry analysis, we also perform hydrodynamic simulations to show that the moment of inertia of the trapped gas indeed follows a universal time-dependence determined jointly by the conformal symmetry and time-dependent scattering length $a(t)$. Our results should also be relevant to the dynamics of other systems that are nearly scale invariant and that are governed by a non-relativistic conformal symmetry.  
\end{abstract}
\maketitle

{\it Introduction.} -- Conformal symmetry~\cite{Hagen72, Niederer72, Henkel94} imposes severe constraints on the dynamics of non-relativistic scale invariant quantum systems, thanks to an overarching SO(2,1) symmetry that generally occurs for quantum systems with dynamical critical exponent $z=2$~\cite{Rosch97, Castin06, Wingate06, Nishida07, Gritsev10, Maki19}. The most prominent example of such systems are atomic gases, where the quantum critical point can be accessed using a Feshbach resonance~\cite{Chin10}. The dynamics of these gases are in fact constrained by conformal symmetry rather than the scale symmetry of the Hamiltonian, independent of the microscopic details of the system.

When an atomic gas deviates from its critical point, the scale invariance of the Hamiltonian is broken, and the interactions can be described by a finite length scale, $a$, that parametrizes the strength of the interactions. Previously studies have focused on the dynamical effects caused by time-independent interactions ~\cite{Maki18, Maki19}. However, since conformal symmetry is dynamical, one can further consider the dynamics of driven quantum critical systems. Indeed, as we shall show in this Letter for the case of a dilute two-component Fermi gas,  with interactions characterized by a time-dependent length scale, $a(t)$, that breaks the conformal symmetry, the long-time dynamics can be either fixed entirely by the conformal symmetry, or in the case when it is not, can be shown to depend on $a(t)$ in a universal fashion. We derive the explicit expression describing the breaking of conformal symmetry based on how $a(t)$ transforms under the conformal transformation.  To give an explicit example, we study how a time-dependent $a(t)$ affects the expansion dynamics and the elliptic flow of Fermi gases near the strongly interacting quantum critical point, and for weak interactions. Both the strongly interacting quantum critical point and the non-interacting point represent scale invariant fixed points (SIFP) in terms of the renormalization group flow \cite{Sachdev}. The dynamics at these SIFPs can be described using conformal symmetry arguments.

It is generally expected that the long-time dynamics of a many-body system would lead to thermalization when the effects of perturbation subsides. In this case, the breaking of conformal symmetry also leads to thermalization~\cite{Maki20}, and in addition imposes its unique constraints on the thermalization process. By tuning the time-dependence of $a(t)$, it turns out that one can not only change the thermalization rate, but also engineer power-law thermalization. This provides a unique route for experimentalists to access the implication conformal symmetry in simple experimental settings, such as the damping of monopole oscillations.



{\it Breaking of Conformal Symmetry with Time-Dependent Interactions.} -- We begin by considering the expansion dynamics of a two-component Fermi gas near the strongly interacting SIFP, or for free particles in $d$-dimension. For specificity, we will focus on the dynamics near the strongly interacting SIFP, and present general results later. Close to the strongly interacting SIFP, the total Hamiltonian can be expanded as~\cite{Maki18, Maki19}:
\begin{equation}
H(t) \approx H_s + \frac{1}{a^{d-2}(t)}C_a,
\label{eq:Hamiltonian}
\end{equation} 

\noindent where $H_s$ is the scale invariant Hamiltonian at the strongly interacting SIFP, and $C_a$ is the contact operator~\cite{Note_Contact,Tan08, Braaten08}. Here $a(t)$ is the time-dependent length scale that parametrizes the interactions. In $d=3$, this length scale is simply the $s$-wave scattering length, and thus we will refer to $a(t)$ as a scattering length. The strongly interacting SIFP is defined when: $1/a^{d-2}(t) = 0$. Eq.~(\ref{eq:Hamiltonian}) is valid to $O(a^{2-d}(t))$, and is a good approximation provided that the rate of change of $a(t)$ is much slower than that set by the range $r_0$ of the two-body potential, $\hbar/mr_0^2$, which is usually very well satisfied in experiments. 

As shown in Ref.~\cite{Maki19}, the dynamics of the density matrix, $\rho(t)$, can be split into a matrix governed by the scale invariant Hamiltonian, $\rho_s(t)$, and a matrix governed by the initial conditions and the breaking of conformal symmetry, $\Gamma(t)$: $\rho(t) \approx \rho_s(t) \Gamma(t)$. The trivial dynamics  of $\rho_s(t)$ are solely controlled by the so-called conformal tower states, which are defined as the eigenstates of the scale invariant gas inside an isotropic harmonic potential,
\begin{equation}
\left[H_s +\omega_0^2 C \right]|n, \ell\rangle = E_n^{\ell} |n, \ell\rangle,
\label{eq:conformal_towers}
\end{equation}
where $C=\frac{1}{2}m\sum_i{\bf r}_i^2$ is the moment of inertia. The dynamics are equivalent to a time-dependent rescaling of the position coordinates, up to a gauge transformation \cite{Maki18, Castin04, Castin06}. In Eq.~(\ref{eq:conformal_towers}),  $\omega_0$ is an arbitrary trapping frequency, and $|n\ell\rangle$ is the $n$th state in the $\ell$th conformal tower with energy $E_n^{\ell}$. The trivial dynamics associated with the conformal towers have previously been used to study a variety of dynamical phenomena \cite{Kagan96, Castin04, Castin06, Minguzzi05, Campo11, Moroz12, Qu16, Deng16, Gharashi16, Kheruntsyan, Deng18, Diao18,Maki18, Maki19, Maki20}. On the other hand, the non-trivial conformal symmetry breaking dynamics contained in $\Gamma(t)$ satisfy the following differential equation:
\begin{align}
\partial_t \Gamma(t) = \frac{i}{a^{d-2}(t)} \left[e^{i H_s t} C_a e^{-i H_s t}, \Gamma(t)\right]
\label{eq:Gamma(t)}
\end{align}
where $\Gamma(0) = \rho_0$ is the initial density matrix.

The effect of the scale symmetry breaking interactions, or equivalently $\Gamma(t)$, can be understood by examining the matrix elements of $\Gamma(t)$ with respect to the conformal tower states to first order in perturbation theory:
\begin{align}
\langle n' \ell' | &\Gamma(t \gg \omega_0^{-1}) | n, \ell\rangle  \approx \langle n' \ell' | \Gamma(0) | n, \ell\rangle \nonumber \\
&+i \int_{\omega_0^{-1}}^t \frac{dt'}{\lambda^2(t')} \left(\frac{\lambda(t')}{a(t')}\right)^{d-2} \langle n'\ell'| \left[\tilde{C}_a, \Gamma(0)\right] | n\ell\rangle
\label{eq:td_pert}
\end{align}
where $\lambda(t) = \sqrt{1+(\omega_0 t)^2}$ and $\langle n'\ell'|\tilde{C_a}|n \ell\rangle =  \exp[i\pi(E_{n'}^{\ell'}-E_n^{\ell})/(2\omega_0)] \langle n'\ell'|C_a|n\ell\rangle $. Eq.~(\ref{eq:td_pert}) states that all the non-trivial dynamics contained in $\Gamma(t)$ must be a function of the effective coupling constant:
\begin{equation}
\mathnormal{g}(t) = \int_{\omega_0^{-1}}^t\frac{dt'}{\lambda^2(t')} \left(\frac{\lambda(t')}{a(t')}\right)^{d-2}
\label{eq:F(t)}
\end{equation}
to first order in perturbation theory. For the case of free space expansion \cite{Maki18}, one can show that this structure persists to higher orders in perturbation theory when $\mathnormal{g}(t)$ diverges in the long limit. As a result, in the long-time limit, $\Gamma(t)$ will be a function of only $\mathnormal{g}(t)$, up to corrections of $O((\omega_0 t)^{-1})$, i.e. $\Gamma(t \gg \omega_0^{-1}) \approx \Gamma(\mathnormal{g}(t))$.

If $\mathnormal{g}(t)$ saturates in the long-time limit, then $\Gamma(t\gg \omega_0^{-1})$ tends to a constant and the dynamics are entirely controlled by $\rho_s(t)$ which in turn is expressed by the time-dependent rescaling of the conformal tower states. In other words, there is an emergent conformal symmetry constraining the dynamics in the long-time limit. Alternatively, if $\mathnormal{g}(t)$ has non-trivial dynamics in the long-time limit, the effects of the conformal symmetry breaking perturbation will be important and there will be no emergent conformal symmetry.

To exemplify this difference, we consider a power law time-dependence for the scattering length:
\begin{equation}
a(t) = a_0 (\eta t)^{\gamma},
\label{eq:a(t)}
\end{equation}
where $\gamma$ is a real number, while $a_0$ and $\eta$ have units of length, and frequency, respectively. For this perturbation, one can evaluate Eq.~(\ref{eq:F(t)}) analytically:
\begin{equation}
\mathnormal{g}(t \gg \omega_0^{-1}) = \left(\frac{\omega_0}{a_0 \eta^{\gamma}}\right)^{d-2}\frac{t^{d-3- \gamma(d-2)}}{d-3 - \gamma(d-2)}  + \mathnormal{g}_0
\label{eq:F_analytical}
\end{equation}
for some constant $g_0$ that describes the short-time physics. $g(t)$ has a non-trivial time dependence in the long-time limit if:
\begin{align}
d-3 &\geq (d-2)\gamma & \text{strong interactions}. \label{eq:con_res}
\end{align}

In three-dimensions, Eq.~(\ref{eq:con_res}) reduces to $\gamma \leq 0$. Therefore, any static perturbation, $\gamma =0$, is marginal, while any perturbation that moves away from the strongly interacting SIFP, $\gamma <0$, is relevant, consistent with previous studies \cite{Maki18,Maki19}. Similarly, in one-dimension we find that a perturbation near the strongly interacting SIFP becomes relevant when $\gamma \geq 2$. In this case, a static perturbation is irrelevant, and one needs to have a scattering length that increases quadratically, $\gamma = 2$, in order to break the conformal symmetry. Therefore, the one-dimensional strongly interacting SIFP is much more resilient against perturbations.

\begin{figure}
\includegraphics[scale=0.65]{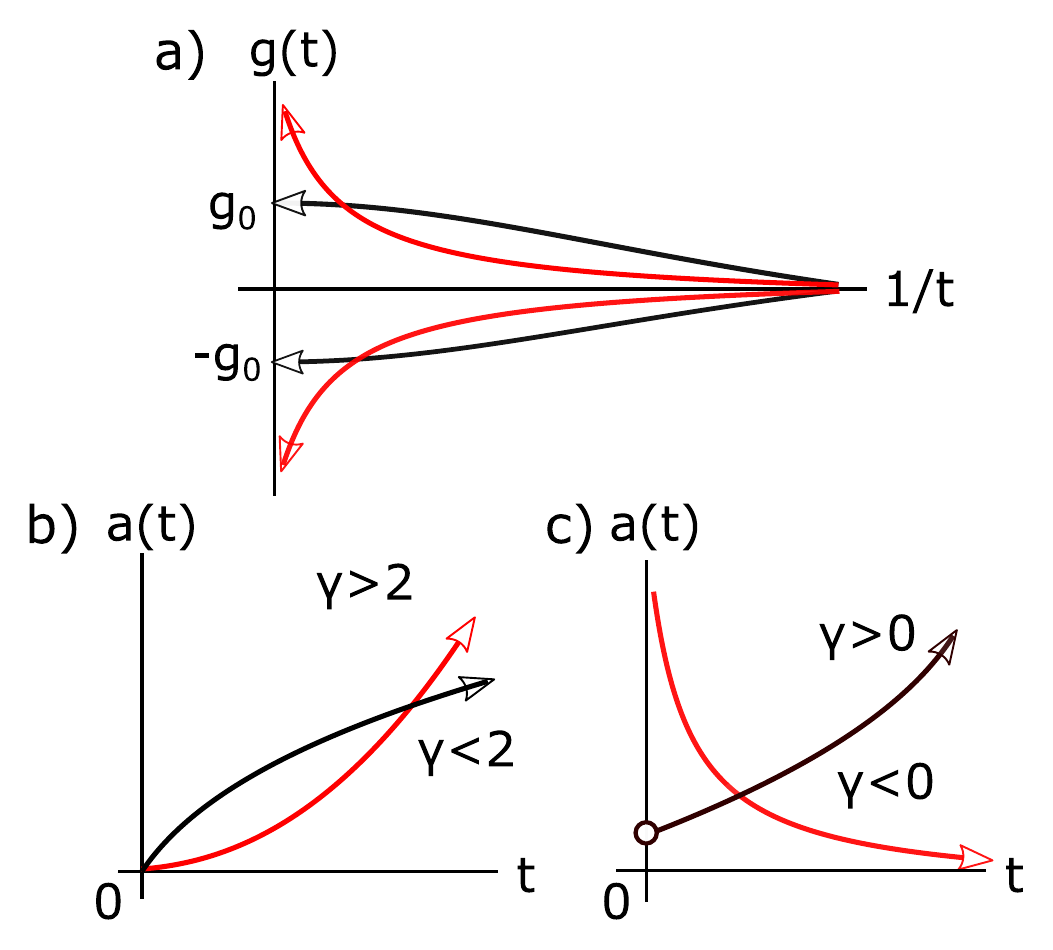}
\caption{a) Schematic of the effective coupling constant, $g(t)$, near a SIFP, $g(t) =0$. The definition of $g(t)$ is given by Eq.~(\ref{eq:F(t)}) near the strongly interacting SIFP, and by Eq.~(\ref{eq:F(t)_weak}) for weak interactions. The black line represents irrelevant symmetry breaking interactions, while the red line denotes relevant ones. The relevancy condition for $\gamma$ is given by Eq.~(\ref{eq:con_res}) near the strongly interacting SIFP, and Eq.~(\ref{eq:con_weak}) for weak interactions. b) and c) depicts the physical time dependence of $a(t)$ for relevant and irrelevant interactions near the strongly and weakly interacting SIFPs in $d=3$, respectively.}
\label{fig:schematic}
\end{figure}

Although Eqs.~(\ref{eq:F(t)}) and (\ref{eq:con_res}) were derived for strongly interacting systems, similar arguments apply for the case of weakly interacting gases \cite{Harmonic_Note}. Expanding the Hamiltonian around the non-interacting SIFP, $a^{d-2}(t) = 0$, one finds the following definition of the effective coupling constant:
\begin{equation}
g(t) = \int_{\omega_0^{-1}}^t \frac{dt'}{\lambda^2(t')} \left(\frac{a(t')}{\lambda(t')}\right)^{d-2},
\label{eq:F(t)_weak}
\end{equation}
and the following criterion for the breaking of conformal symmetry:
\begin{align}
d-1 &\leq (d-2)\gamma & \text{weak interactions.}
\label{eq:con_weak}  
\end{align}
The relevancy of $a(t)$ in the weak coupling limit reads $\gamma \geq 2$ in three-dimension while for one-dimension: $\gamma \leq 0$. The three- (one-) dimensional criterion near the strongly interacting SIFP is equivalent to the one- (three-) dimensional criterion for weak interactions. In Fig.~(\ref{fig:schematic}), we show a schematic for how the effective coupling constant, $g(t)$, changes with time, and the associated time-dependence of the scattering length, $a(t)$, near both the strongly interacting and non-interacting SIFPs in three dimensions. The relevancy of a perturbation is equivalent to whether the effective coupling constant, $g(t)$, increases or decreases with time.

In the case of two-dimensions, there is a single non-interacting SIFP. Eq.~(\ref{eq:con_weak}) then states that any scattering length with a power-law time dependence is irrelevant to the expansion dynamics in two-dimensions. Therefore one needs a stronger time-dependence for $\mathnormal{g}(t)$, to exhibit non-trivial dynamics. This conclusion is consistent with the two-body interaction depending on the scattering length logarithmically, i.e. the quantum anomaly \cite{Olshanii10, Vogt12, Hofmann12, Gao12,Murthy19}. 

{\it Consequences of Broken Conformal Symmetry on Eliptic Flow}. -- The conclusions of the previous section are independent of the initial state of the system, and depend only on the conformal SO(2,1) symmetry in the free space expansion. To elucidate the physical consequences of this symmetry, let us consider the effects of a time-dependent perturbation on the expansion dynamics of three-dimensional Fermi gases prepared in an anisotropic trap, i.e. eliptic flow. This was studied experimentally both at resonance and in the presence of a static scale breaking perturbation Ref.~\cite{Cao11, Elliott14}. In order to parametrize the eliptic flow, we use the moments of inertia in the directions, $i = x,y,z$, defined as:
\begin{equation}
\langle r_i^2 \rangle(t) = \frac{1}{N} \int d{\bf r} r_i^2 n(r,t),
\end{equation}
where $n(r,t)$ is the time-dependent density of the gas, or equivalently the diagonal elements of the one-body density matrix in position space. From conformal symmetry arguments, the moment of inertia in the long-time limit has the form:
\begin{equation}
\langle r_i^2 \rangle(t \gg \omega_i^{-1}) \approx \left[A_i + \frac{B_i}{\omega_i t} + \frac{D_i}{\left(\omega_i t\right)^2} + E_i \mathnormal{g}(t)\right]t^2,
\label{eq:moi}
\end{equation} 
where terms involving $A_i$, $B_i$, and $D_i$ are guaranteed by conformal symmetry, and their numerical value depends on the initial conditions \cite{Initial_Condition_Note} and the short-time dynamics. On the other hand, the scale breaking term depends linearly on the effective coupling constant, $\mathnormal{g}(t)$, defined in Eq.~(\ref{eq:F(t)}), with a proportionality constant $E_i$.

Depending on the time-dependence of $\mathnormal{g}(t)$, one can change the asymptotic behavior of the moment of inertia. If the perturbation is relevant, the scale breaking term proportional to $\mathnormal{g}(t)$ becomes parametrically larger than the conformal symmetric terms in the long-time limit. A similar situation can also occur for $0 < \gamma  <1$. In this case the perturbation is irrelevant to the long-time dynamics, but the leading correction to the conformal physics is now given by the scale breaking term: $\mathnormal{g}(t) \propto t^{-\gamma}$.

Eq.~(\ref{eq:moi}) also has consequences for the aspect ratio, which is the ratio between the moments of inertia in two given directions, say $x$ and $y$. If the breaking of scale invariance provides the leading correction to the dynamics, the aspect ratio will have the perturbative form:
\begin{equation}
\lim_{t\to\infty}\frac{\langle r^2_x\rangle(t)}{\langle r^2_y\rangle(t)} \approx \frac{A_x}{A_y}\left[1 + \left(\frac{E_x}{A_x}- \frac{E_y}{A_y}\right)\mathnormal{g}(t)\right].\label{eq:aspectratio}
\end{equation}
Conformal symmetry requires that the aspect ratio saturates to a constant with a correction of $O((\omega_0 t)^{-1})$ in the long-time limit. However, for irrelevant perturbations with $0< \gamma <1$, the aspect ratio still approaches a constant in the long time limit but with a different time dependence. If the interaction is relevant though, $\gamma \leq 0$, the aspect ratio will not saturate due to the non-trivial time dependence induced by the conformal breaking interactions. Both these situations can be tested in atomic gas experiments.

Eqs.~(\ref{eq:moi},\ref{eq:aspectratio}) constitute one of the main results of this Letter. By tuning the time-dependences of the scattering length $a(t)$, it is possible to change the asymptotic behavior of the expansion dynamics, as described by $\mathnormal{g}(t)$. It is to be emphasised that the simple dependence on $\mathnormal{g}(t)$ reflects not only the time-dependence of $a(t)$, but also the conformal symmetry that is present at the SIFP, as is evident in the derivation of Eq.~(\ref{eq:F(t)}). Thus an experimental confirmation of the Eqs.~(\ref{eq:moi},\ref{eq:aspectratio}) would constitute an indirect experimental verification of the existence of conformal tower states. 

{\it Hydrodynamics with Time-Dependent Interactions.} -- Let us now compare this approach to the standard hydrodynamics~\cite{Elliott14, Fujii18,Landau} when it is applicable. To this end we examine the expansion dynamics of a trapped Fermi gas near the strongly interacting SIFP with time-dependent interactions, when subject to a quench of the harmonic trapping potential: $\omega_0 \to \omega_f$, where $\omega_f \ll \omega_0$. For simplicity, we will assume that both the initial and final trapping potential are isotropic. Thus the only source of broken conformal symmetry is due to the changing interaction. To describe the expansion, we will focus on the total moment of inertia, and assume it satisfies the following scaling ansatz:
\begin{equation}
\langle r^2 \rangle(t) = \sum_{i = x,y,z} \langle r_i^2 \rangle(t) = \lambda_a^2(t) \langle r^2 \rangle(0).\label{hydror2}
\end{equation}
From standard hydrodynamic arguments~\cite{Elliott14, Fujii18,Landau} and by performing an expansion near the strongly interacting SIFP, one can obtain the following differential equation:
\begin{align}
&\frac{d^2\lambda_a^2(t)}{dt^2} = 2(\omega_0^2 + \omega_f^2) - 4\omega_f^2 \lambda_a^2(t) \nonumber \\
&+ \tilde{C}_a \left[\frac{1}{\lambda_a^{4-d}(t)} \left(\frac{1}{\tilde{a}(t)}\right)^{d-2} - 1 \right] \nonumber \\
&+  2\tilde{C}_a \int_0^t dt' \ \left(\frac{1}{\tilde{a}(t')}\right)^{d-1} \frac{1}{\lambda_a^{4-d}(t)} \frac{d\tilde{a}(t')}{dt'} \nonumber \\
&- \tilde{\zeta} \left(\frac{\lambda_a^2(t)}{\tilde{a}^2(t)}\right)^{d-2} \left( \frac{1}{\lambda_a^2(t)}\frac{d\lambda_a^2(t)}{dt} - \frac{2}{\tilde{a}(t)} \frac{d\tilde{a}(t)}{dt}\right)
\label{eq:eom_final}
\end{align}
where we have defined: 
\begin{align}
\tilde{a}(t) &= \frac{a(t)}{a(0)}, &
\tilde{C}_a &=  \frac{2\langle C_a \rangle(0)}{\langle r^2 \rangle(0) a^{d-2}(0)}, &
\tilde{\zeta} &= d^2 \int \frac{d{\bf r}}{N} \frac{\zeta({\bf r},0)}{ \langle r^2 \rangle(0)}, \nonumber \\
\label{eq:eom_constants}
\end{align}

\noindent and $\zeta({\bf r},t)$ is the local bulk viscosity which depends quadratically on the inverse scattering length, $1/a(t)^2$, and has an approximate scaling form near the SIFP \cite{Schaefer13, Enss19, Nishida19, Hofmann20}. Finally we note the following initial conditions: $\lambda_a(0) = 1$ and $\dot{\lambda}_a(0) = 0$, and for numerical simulations, we will initialize our system slightly away from the SIFP.

First, let us consider the expansion into free space ($\omega_f =0$). At the strongly interacting SIFP, $a(t)=\infty$, the solution to Eq.~(\ref{eq:eom_final}) is $\lambda^2(t) = 1+(\omega_0 t)^2$, the same as that based on conformal symmetry. However, when there is a finite time dependent $a(t)$, the conformal symmetry is broken which leads to corrections in $\lambda^2(t)$ that are described by $\mathnormal{g}(t)$ given in Eq.~(\ref{eq:moi}) [see also Eq.~(\ref{hydror2})]. This can also be seen from the hydrodynamic formalism. Let the solution of Eq.~(\ref{eq:eom_final}) be given by $\lambda_a^2(t)$ and consider its difference from $\lambda^2(t)$:  $\delta \lambda^2(t) \equiv (\lambda_a^2(t) - \lambda^2(t))/\lambda^2(t)$. In the long time limit and for weak conformal symmetry breaking, one expects that $\delta \lambda^2(t)$ to be simply proportional to $\mathnormal{g}(t)$. 

\begin{figure}
\includegraphics[scale=0.6]{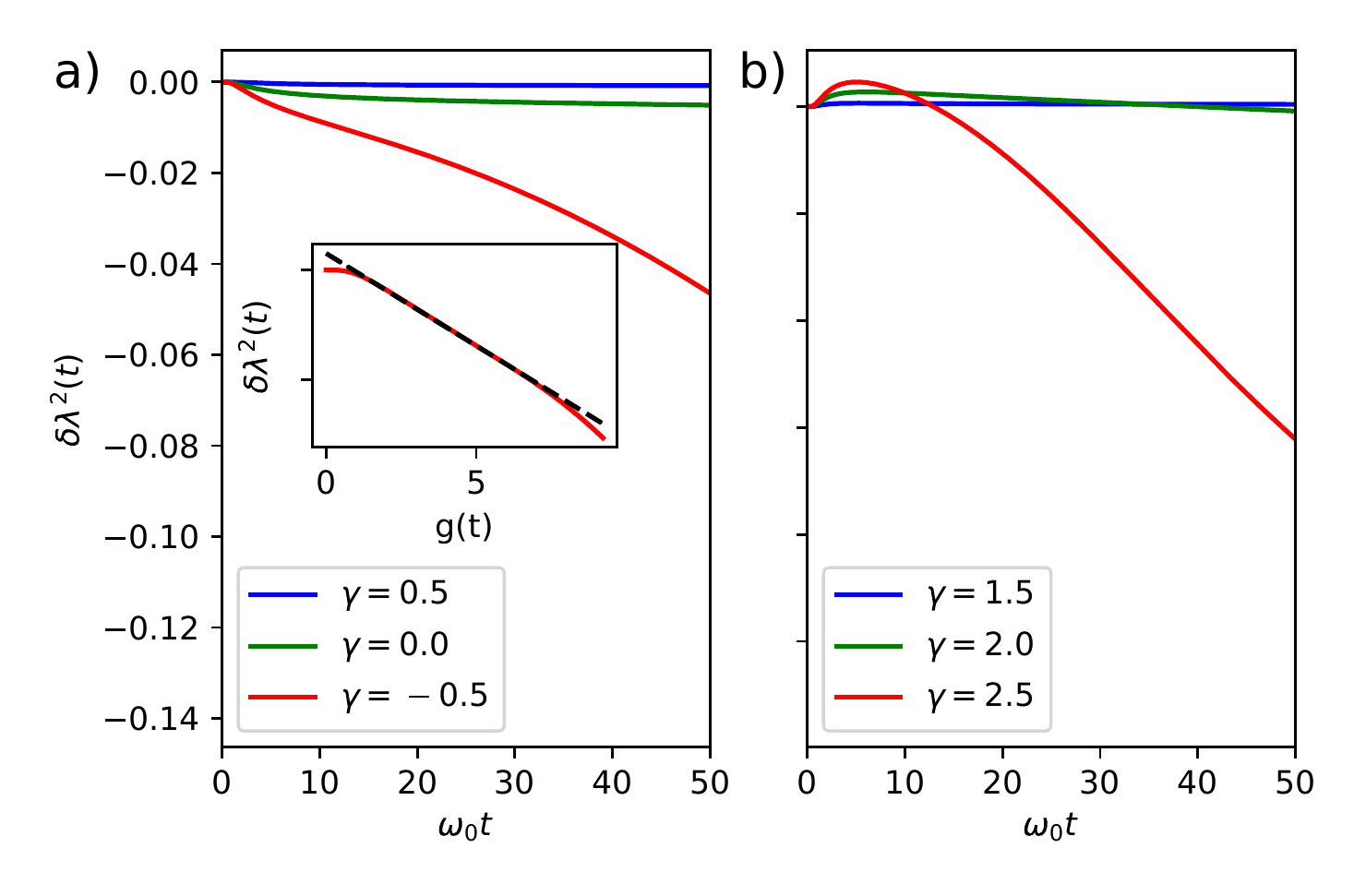}
\caption{Solutions to the hydrodynamic equations [Eq.~(\ref{eq:eom_final})] in free space with a time-dependent scattering length [Eq.~(\ref{eq:a(t)})] near the strongly interacting SIFP for a) 3D and b) 1D. Here we set $\eta = 1.5 \omega_0$. For relevant symmetry breaking perturbations, i.e. $\gamma=-0.5$ in 3D and $\gamma=2.5$ in 1D, the deviation from the conformal solution becomes significant and is accurately described by Eq.~(\ref{eq:moi}). The inset in a) shows the linear relation between $\delta \lambda^2(t)$ and $\mathnormal{g}(t)$, which holds for approximately $0 < \omega_0 t < 30$. For irrelevant symmetry breaking interactions, $\gamma=0.5$ in 3D and $\gamma=2$ in 1D, the hydrodynamic simulation closely tracks the conformal solution in the long time limit.} 
\label{fig:free_space_expansion}
\end{figure}

In Fig.~(\ref{fig:free_space_expansion}), we present $\delta \lambda^2(t)$ in both $d=3$ [Fig.(\ref{fig:free_space_expansion}a)] and $d=1$  [Fig.(\ref{fig:free_space_expansion}b)], and for various value of $\gamma$ in Eq.~(\ref{eq:a(t)}). As one can see in both cases the differences becomes substantially larger when $\gamma$ satisfies the relevancy condition, Eqs.~(\ref{eq:con_res}). On the other hand, if the symmetry breaking interaction is irrelevant, then the long time dynamics closely track the conformal invariant solution. In addition, as we show in the inset, $\delta \lambda^2(t)$ is also proportional to $g(t)$ [see Eq.~(\ref{eq:moi})] for a substantial time window that extends to a time $t$ about $\omega_0t\approx 30$, which would allow enough time for observation experimentally. The deviation from linearity at even longer times is due to the fact that in the hydrodynamic theory we have included the bulk viscosity: $\zeta\propto 1/a^{2(d-2)}(t)$. We note that this approximate form of the bulk viscosity is only valid around the strongly interacting SIFP and becomes a bad approximation when $a(t)$ substantially deviates from its critical value. 

Next, let us consider the case of a quenched harmonic potential: $\omega_0 \to \omega_f$. At the SIFP, both $\tilde{C}_a$ and $\tilde{\zeta}$ vanish. Eq.~(\ref{eq:eom_final}) then predicts undamped oscillations at exactly $2\omega_f$, identical to the scale invariant solution \cite{Harmonic_Note}. Such undamped oscillations are indicative of reversible, entropy conserving dynamics~\cite{Maki20}, and have been observed near the non-interacting SIFP \cite{Lobser15}. When the scale invariance is broken, there will be irreversible damping due to the finite bulk viscosity, and hence entropy production. Since the initial state is highly excited compared to the ground state of the final trap~\cite{Maki20}, and since the motion of the gas evolves at a much slower rate than the relaxational physics set by the Boltzmann time scale, it is natural that the decay of these oscillations is also indicative of thermalization.

\begin{figure}
\includegraphics[scale=0.6]{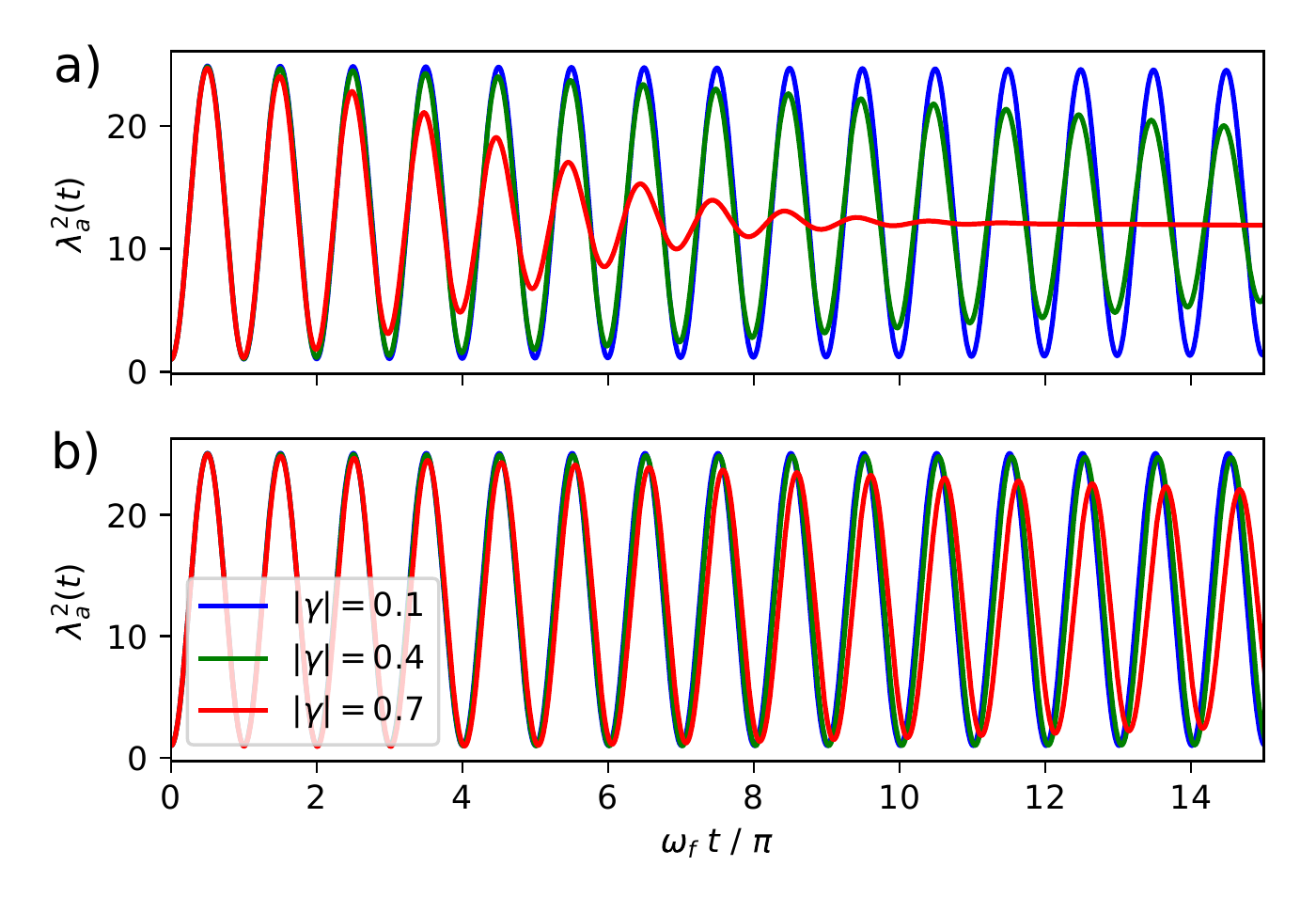}
\caption{Numerical solution the hydrodynamic equations [Eq.~(\ref{eq:eom_final})] when one moves away from the strongly interacting SIFP for a) $d=3$ and b) $d=1$. To compare these two situations, the symmetry breaking interactions in the Hamiltonian, Eq.~(\ref{eq:Hamiltonian}), are equivalent in $d=3$ and $d=1$.  I.e. we use the same values of the coupling constants, $\eta = \omega_0$, $|\tilde{C}_a| = 0.01$, $\tilde{\zeta} = \tilde{C}_a^2$ for both $d=1,3$ (For repulsive interactions $\tilde{C}_a >0$ for $d=3$ and $\tilde{C}_a<0$ for $d=1$), and the same value of $|\gamma|$ ($\gamma <0$ for $d=3$, and $\gamma >0$ for $d=1$).}
\label{fig:damped_hydro}
\end{figure}

In Fig.~(\ref{fig:damped_hydro}), we show the numerical solutions of Eq.~(\ref{eq:eom_final}) near the strongly interacting SIFP in three-dimensions [Fig.~(\ref{fig:damped_hydro}a)] and one-dimension [Fig.~(\ref{fig:damped_hydro}b)], as one moves away from the SIFP. For an equivalent change in the Hamiltonian, Eq.~(\ref{eq:Hamiltonian}), the one-dimensional Fermi gas is more stable against conformal symmetry breaking than its three-dimensional counterpart. This is exactly what we expect based on the relevancy criterion given in Eq.~(\ref{eq:con_res}). 

The damping of the oscillation amplitude, $\lambda_a^2(t)$, can be described by the following phenomenological equation: $\lambda_a^2(t)=\lambda_a^2(0)\exp[-\Gamma_d(t) t]$, where the effective damping rate, $\Gamma_d(t)$, is given by:
\begin{equation}
\frac{1}{\Gamma_d(t)} \approx \frac{2B}{\omega_0}\frac{\tilde{a}^{2(d-2)}(t)}{\tilde{\zeta}},
\label{eq:Gamma_T}
\end{equation}
for some constant $B$ that depends on the details of the time-dependence of $a(t)$. 

\begin{figure}
\includegraphics[scale=0.6]{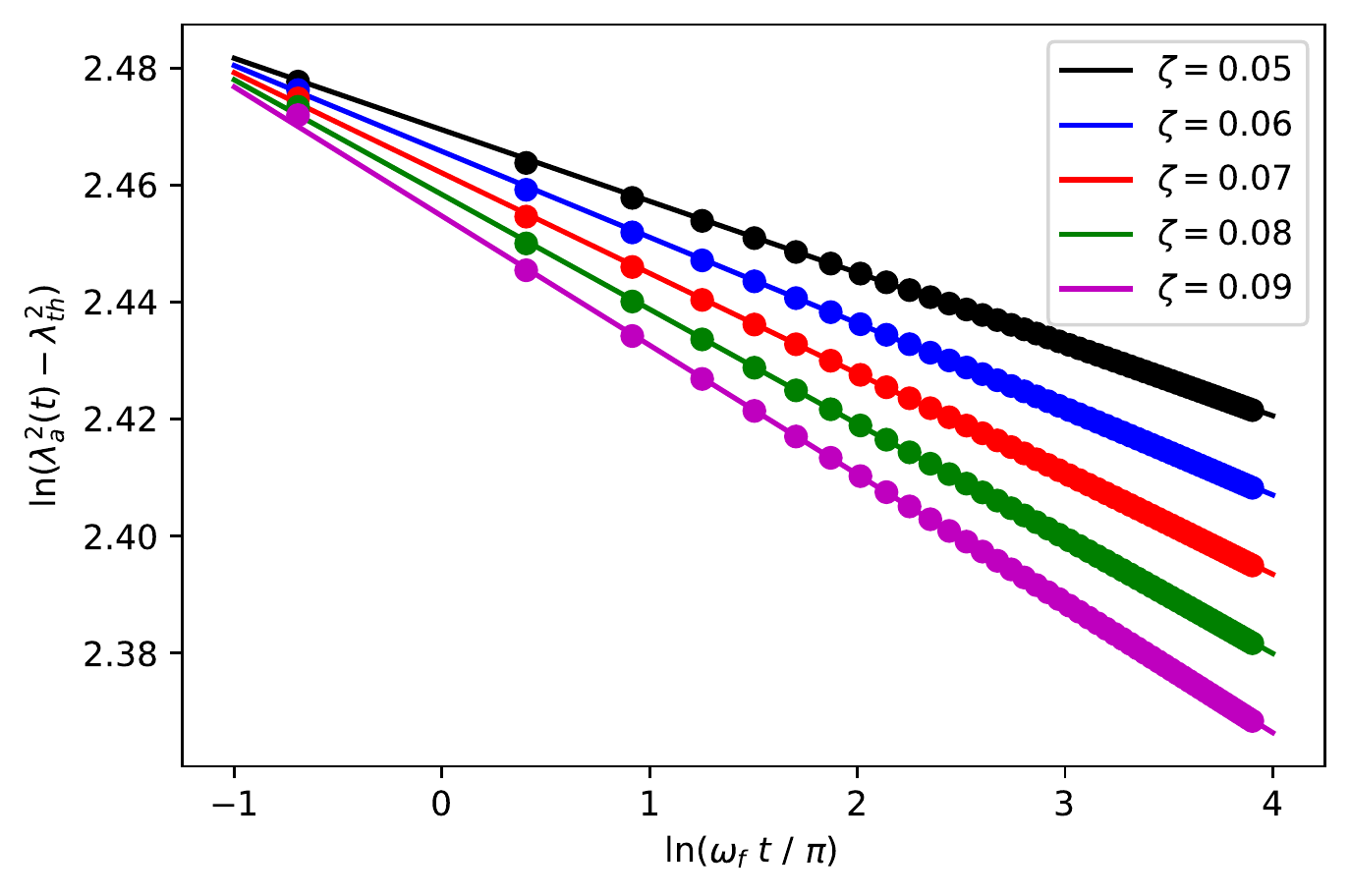}
\caption{Power-law damping of the peak height with respect to the final thermalized value, $\lambda^2_{\rm th}=(\omega_0^2 + \omega_f^2)/(2\omega_0^2)$, in the oscillations of a Fermi gas near the strongly interacting SIFP, as a function of the bulk-viscosity $\tilde{\zeta}$, when $d=3$. Here we set: $\omega_f = \omega_0/5$, $\tilde{C}_a = 0.05$, and $\eta = 2 \omega_0$. The time-dependent scattering length is given by Eq.~(\ref{eq:proposed_a}).}
\label{fig:powerlaw_damping_fit}
\end{figure}

Eq.~(\ref{eq:Gamma_T}) suggests that equilibration/thermalization can be sped up or slowed down by changing the time dependence of the scattering length. Consider the following time-dependent scattering length:
\begin{equation}
\tilde{a}^{d-2}(t) =\left[\frac{1}{ \sqrt{1+ (\eta t)^2}}\right]^{-1/2}.
\label{eq:proposed_a}
\end{equation}
In the long-time limit, $\tilde{a}^{d-2}(t) \propto (\eta t)$. In this case, the damping of the local maxima of the moment of inertia follows a power law behaviour:
\begin{align}
\lambda^2_a(t) &\propto \frac{\omega_0^2+\omega_f^2}{2\omega_0^2}\left(1 + \frac{A}{(\eta t)^{\alpha}}\right) ,& \alpha &= \frac{\tilde{\zeta}}{2\eta}
\label{eq:lambda_sol}
\end{align}
for some constant $A$. In the limit $t\to\infty$, the system thermalizes to a universal value: $\lambda_a^2(t\to\infty)=\lambda^2_{\rm th}=(\omega_0^2 + \omega_f^2)/(2\omega_0^2)$. To test this hypothesis, we fit the local maxima of the oscillations to a power law decay, and find results consistent with Eq.~(\ref{eq:lambda_sol}). The power law decay is shown in Fig.~(\ref{fig:powerlaw_damping_fit}), and as one can see, is very accurate at describing the damping physics.

{\it Conclusions} -- In this letter, we provided a useful criterion for understanding whether a time-dependent scale breaking perturbation is relevant to the long-time dynamics. The time dependence of the perturbation can be used to enhance or diminish the signature of broken conformal invariance, which is useful for experiments. One unique way of using the time-dependent scattering length is to engineer power law thermalization in trapped Fermi gases. Although we focused on the application to atomic gases, we stress that these results extend to other quantum systems with dynamical exponent $z=2$, such as the Lifshitz transition in solid state materials \cite{Lifshitz60}.

\paragraph{Acknowledgements.}
JM and SZ are supported by the Research Grants Council of the Hong Kong Special Administrative Region, China (General research fund, HKU 17304719, 17304820 and collaborative research fund C6026-16W and C6005-17G), and the Croucher Foundation under the Croucher Innovation Award. F.Z. was in part supported by the Canadian Institute for Advanced Research.


\begin{thebibliography}{References}
\bibitem{Note_Contact} In previous works we the breaking of scale symmetry was parametrized by the so-called $V$-matrix. This matrix is identical to the contact matrix. To make this connection more transparent, we use this new notation.
\bibitem{Harmonic_Note} Note that this formalism can also be extended to the motion of scale invariant gases in time-dependent traps. In this case, $\lambda(t)$ in Eq.~(\ref{eq:F(t)}) is replaced by the solution to the differential equation \cite{Castin04,Castin06}: $\omega_0^2 = \ddot{\lambda}(t) \lambda^3(t) + \omega^2(t) \lambda^4(t)$, with $\lambda(0) = 1$ and $\dot{\lambda}(0) = 0$.
\bibitem{Initial_Condition_Note} For isotropic systems, $B = 0$. Therefore, when $\gamma <2$ the leading correction to the dynamics of the moment of inertia is due to the deviation from the SIFP.
\end{thebibliography}
\end{document}